\documentclass[journal]{IEEEtran}
\usepackage{cite}

\ifCLASSINFOpdf

\else

\fi

\usepackage{amsmath}
\usepackage{amsfonts}
\usepackage{array,multirow}
\usepackage[makeroom]{cancel}
\usepackage{graphicx}
\usepackage{makecell}
\usepackage{array}
\usepackage{xcolor}
\usepackage{siunitx}

\begin{document}
	
	\raggedbottom
	
	\newcounter{mytempeqncnt}
	
	\title{Power Consumption and Energy-Efficiency for In-Band Full-Duplex Wireless Systems}
	
	\author{Murad~Murad,~\IEEEmembership{Student Member,~IEEE,} and~Ahmed~M.~Eltawil,~\IEEEmembership{Senior Member,~IEEE}
		\thanks{The authors are with the Department of Electrical Engineering and Computer
			Science at the University of California, Irvine, CA 92697 USA (e-mail:
			mmurad@uci.edu; aeltawil@uci.edu).}
		\thanks{Manuscript received April 22 2019; revised April 22 2019.}
		\thanks{This work has been submitted to the IEEE for possible publication.
			Copyright may be transferred without notice, after which this version may
			no longer be accessible.}}
	
	\markboth{IEEE Wireless Communications Letters,~Vol.~X, No.~X, April~2019}%
	{Murad and Eltawil: Power Consumption and Energy-Efficiency for In-Band Full-Duplex Wireless Systems}
	
	\maketitle
	
	\begin{abstract}
		This paper presents an analytical model of power consumption for In-Band Full-Duplex (IBFD) Wireless Local-Area Networks (WLANs). Energy-efficiency is compared for both Half-Duplex (HD) and IBFD networks. The presented analytical model closely matches the results generated by simulation. For a given traffic scenario, IBFD systems exhibit higher power consumption, however at an improved energy efficiency when compared to equivalent HD WLANs. 
	\end{abstract}
	
	
	\IEEEpeerreviewmaketitle
	
	\section{Introduction}
	\IEEEPARstart{I}{ncreasing} energy-efficiency in IEEE 802.11 Wireless Local-Area Networks (WLANs) has become a priority due to the vast deployment of WiFi networks over recent years \cite{Tsao10}. In this paper, an analytical model is presented to quantify power consumption and energy-efficiency for In-Band Full-Duplex (IBFD) WLANs. Unlike Half-Duplex (HD) communications (i.e. Time-Division Duplexing or Frequency-Division Duplexing), IBFD techniques allow two wireless nodes to transmit and receive simultaneously on one frequency band (details can be found in \cite{Song17}). While IBFD is a promising technique to improve several metrics in wireless networks, the effect on power consumption has not been sufficiently addressed. This paper presents a mathematical model for power consumption in IBFD WLANs, and the model is confirmed by simulation.
	
	\section{System Model}
	This paper assumes an infrastructure WLAN with an Access Point (AP) and $n-1$ associated client stations (STAs) adopting basic IEEE 802.11 Distributed Coordination Function (DCF) standard with one channel. Total frame loss happens when collisions occur. No errors at the PHY layer take place. All wireless nodes can detect one another with no hidden terminals. Each node always has a frame to transmit. The AP always has a load of the maximum MAC Protocol Data Unit (MPDU$_{\text{max}}$). All STAs have equal Symmetry Ratio (SR) values
	as defined in \cite{Murad17}. If the traffic load is designated as ($L$), then SR is the ratio of the uplink to the downlink as follows
	\begin{equation}
		\rho \overset{\Delta}{=} \frac{L_{UL}}{L_{DL}}.
	\end{equation}
	PHY and MAC parameters are set according to the latest IEEE 802.11ac release \cite{80211ac} as indicated in Table \ref{murad.t1}.
	\begin{table} [!t]
		\caption{IEEE 802.11ac PHY and MAC Parameters.}
		\label{murad.t1}
		\centering
		\begin{tabular}{l|r}
			Parameter & Value \\\hline
			Channel bandwidth & 80 MHz \\
			Spatial streams & 2$\times$2 MIMO\\
			PHY header duration & 44 $\mu$s \\
			Transmission rate & 234 Mbps\\
			Basic rate & 24 Mbps\\
			MAC header size & 36 bytes \\
			FCS size & 4 bytes \\
			ACK size & 14 bytes \\
			MPDU$_{\text{max}}$  size & 7,991 bytes\\
			Slot duration ($\sigma$) & 9 $\mu$s \\
			SIFS duration & 16 $\mu$s \\
			DIFS duration & 34 $\mu$s \\
			CW$_{\text{min}}$ & 16 \\
			CW$_{\text{max}}$ & 1024
		\end{tabular}
	\end{table}
	\section{HD IEEE 802.11 Power Consumption Model}
	Power consumption is based on the classical definition of power 
	\begin{equation}
		\text{Power} \overset{\Delta}{=}  \frac{\text{Energy}}{\text{Time}}.
		\label{power}
	\end{equation}
	
	As presented in \cite{Ergen07}, the energy consumed by a node in an HD WLAN depends on the state of the node. There are six mutually exclusive states a node can be in. The energy consumption $(\mathcal{E})$ in terms of power consumption $(\omega)$ and the probability for each state are given as follows
	\begin{enumerate}
		\item Idle (d) state
		\begin{equation}
			\mathcal{E}_{\text{d}} = \omega_{\text{d}}\sigma
		\end{equation}
		\begin{equation}
			Pr(\text{d}) = (1-\tau)^n
		\end{equation}
		
		\item Successful transmission (S-TX) state
		\begin{equation}
			\hspace{-0.50cm}\mathcal{E}_{\text{\tiny S-TX}}= \omega_{\text{\tiny TX+CTRL}}\text{DATA}_{\text{\tiny TX}}+ \omega_{\text{d}} (\text{DIFS+SIFS})+ \omega_{\text{\tiny RX+CTRL}} \text{ACK}\hspace{-0.09cm}
		\end{equation}
		\begin{equation}
			Pr(\text{S-TX}) = \tau (1-p)
		\end{equation}
		
		\item Successful reception (S-RX) state
		\begin{equation}
			\hspace{-0.50cm}\mathcal{E}_{\text{\tiny S-RX}} = \omega_{\text{\tiny RX+CTRL}} \text{DATA}_{\text{\tiny RX}} + \omega_{\text{d}}  (\text{DIFS+SIFS}) + \omega_{\text{\tiny TX+CTRL}} \text{ACK}\hspace{-0.11cm}
		\end{equation}
		\begin{equation}
			Pr(\text{S-RX}) = \tau (1-\tau)^{n-1}
		\end{equation}
		
		\item Successful overhearing (S-$\overline{\text{RX}}$) state
		\begin{equation}
			\mathcal{E}_{\text{\tiny S-$\overline{\text{RX}}$}} = \omega_{\text{\tiny RX+CTRL}} (\text{DATA}_{\text{\tiny RX}}+\text{ACK}) + \omega_{\text{d}} (\text{DIFS+SIFS})
		\end{equation}
		\begin{equation}
			Pr(\text{S-RX}) = (n-2)\tau (1-\tau)^{n-1}
		\end{equation}
		
		\item Transmitting during a collision (C-TX) state
		\begin{equation}
			\mathcal{E}_{\text{\tiny C-TX}} = \omega_{\text{\tiny TX+CTRL}} \text{DATA}_{\text{\tiny TX}} + \omega_{\text{d}} (\text{DIFS+SIFS+ACK})
		\end{equation}
		\begin{equation}
			Pr(\text{C-TX}) = \tau p
		\end{equation}
		
		\item Overhearing a collision (C-$\overline{\text{RX}}$) state
		\begin{equation}
			\mathcal{E}_{\text{\tiny C-RX}} = \omega_{\text{\tiny RX+CTRL}} \text{DATA}_{\text{\tiny RX}} + \omega_{\text{d}} (\text{DIFS+SIFS+ACK})
		\end{equation}
		\begin{equation}
			\hspace{-0.7cm}Pr(\text{C-$\overline{\text{RX}}$}) = (1-\tau)[1-(1-\tau)^{n-1}-(n-1)\tau(1-\tau)^{n-2}]\hspace{-0.2cm}
		\end{equation}
	\end{enumerate}
	where $\tau$ is the probability of transmission, $p$ is the conditional collision probability, and $n$ is the number of nodes (details are in \cite{Bianchi00}).
	
	The expected value of consumed energy by a node can be expressed in terms of the energy consumption and probability of each state as
	\begin{equation}
		\mathbb{E}[\text{energy}] = \sum_{i=1}^{6} (\text{energy in state \textit{i}}) \cdot Pr(\text{state \textit{i}}).
		\label{E[energy]} 
	\end{equation}
	
	Finally, the average power consumption of a node is given by rewriting (\ref{power}) as  
	\begin{equation}
		\begin{split}
			\text{Power} &= \frac{\mathbb{E}[\text{energy}]}{\mathbb{E}[\text{time duration}]}\\
			\\
			&= \frac{\sum_{i=1}^{6}(\text{energy in state \textit{i}}) \cdot Pr(\text{state \textit{i}})}{(1-P_{tr})\sigma+P_{tr} P_{s} T_{s} + P_{tr}(1-P_{s})T_{c}}
		\end{split}
	\end{equation}
	where the probability that there is at least a transmission $(P_{tr})$, the probability of a successful transmission $(P_s)$, the expected time of a successful transmission $(T_s)$, and the expected time of a collision $(T_c)$ are are given in \cite{Bianchi00}.
	
	\section{Analysis for IBFD WLAN Power Consumption}
	A similar approach to the HD case is adopted here for an  IBFD WLAN. Several considerations must be taken into account. First, the AP properties in an IBFD system are different from those of STAs. Second, energy consumption for Self-Interference Cancellation (SIC) to enable IBFD communications must be added. Third, IBFD mechanisms must be factored into the expressions for each state.
	\subsection{The AP in an infrastructure IBFD WLAN}
	For the AP, there are 3 states as follows
	\begin{enumerate}
		\item Idle (AP-d) state
		\item Successful transmission/reception (AP-S-TXRX) state
		\item Transmitting/receiving a collision (AP-C-TXRX) state
	\end{enumerate}
	\subsection{An STA in an infrastructure IBFD-WLAN}
	For each STA, there are 5 states as follows
	\begin{enumerate}
		\item Idle (STA-d) state    
		\item Successful transmission/reception (STA-S-TXRX) state
		\item Successful overhearing (STA-S-$\overline{\text{RX}}$) state
		\item Transmitting/receiving a collision (STA-C-TXRX) state
		\item Overhearing a collision (STA-C-$\overline{\text{RX}}$) state
	\end{enumerate}
	The consumed energy and the probability of each state for the AP and an STA in an IBFD WLAN are given below in (17) through (32). Detailed expressions for $\tau_{_{AP}}$, $\tau_{_{STA}}$, $P_{tr}$, $P_s$, $T_s$, and $T_c$ in an IBFD WLAN can be found in \cite{murad19}.
	\begin{table} [!b]
		\caption{Power Consumption Values.}
		\label{murad.t2}
		\centering
		\begin{tabular}{l|c}
			Power Category & Value \\\hline
			Transmitter ($\omega_{\text{\tiny TX}}$) & 2.6883 W\\
			Receiver ($\omega_{\text{\tiny RX}}$) & 1.5900 W\\
			Idel State ($\omega_{\text{d}}$) & 0.9484 W\\
			Control Circuit ($\omega_{\text{\tiny CTRL}}$) & 0.3000 W\\
			Self-Interference Cancellation ($\omega_{\text{\tiny SIC}}$) & 0.0650 W
		\end{tabular}
	\end{table}
	\begin{figure}[!b]
		\centering
		\includegraphics[width=\columnwidth,trim=1.4in 3.3in 1.7in 3.5in,clip=true]
		{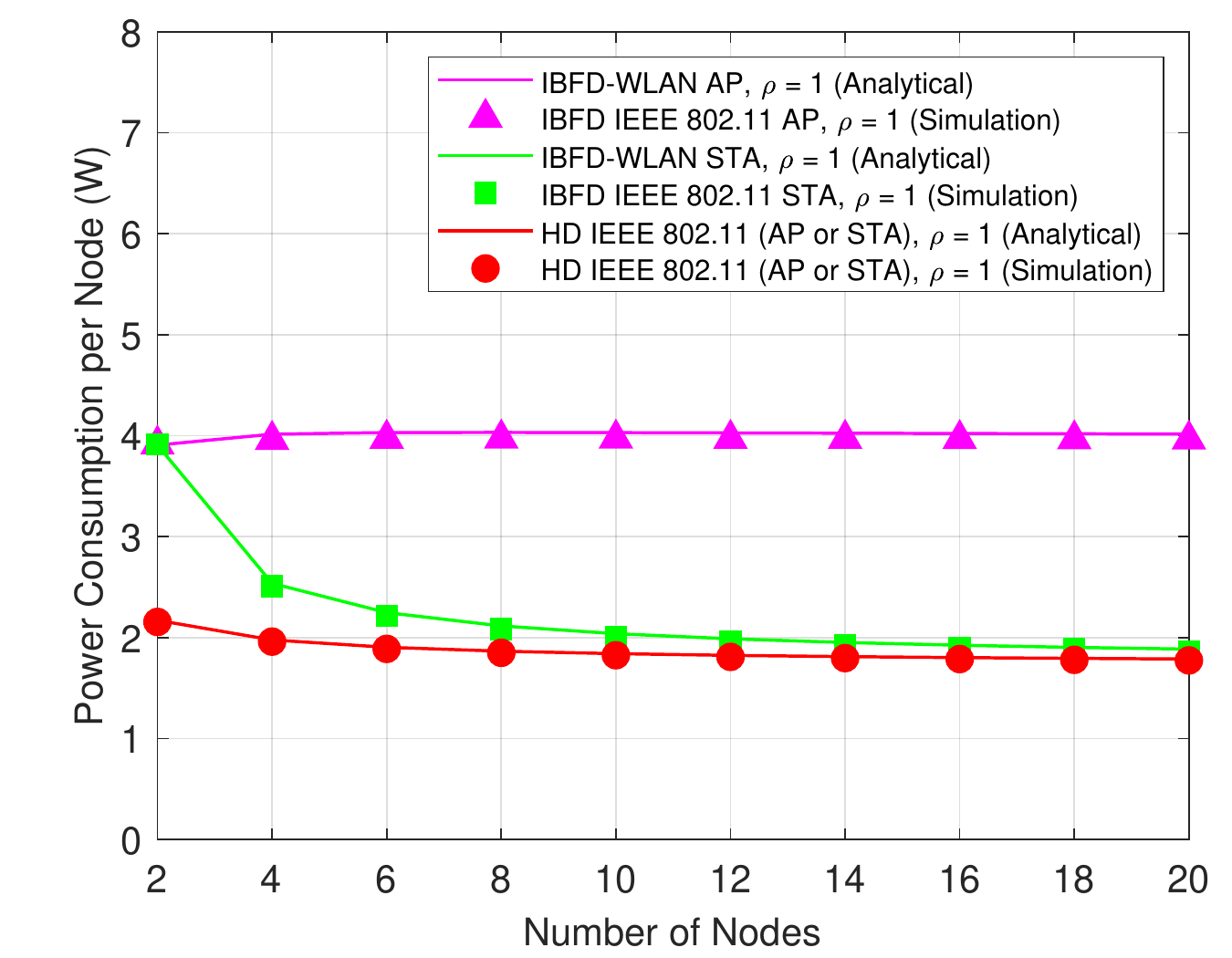}
		\caption{Power consumption per node in HD and IBFD WLANs when $\rho = 1$.}
		\label{murad1}
	\end{figure}
	\begin{figure*}[!t]
		\textit{Energy Consumption for the AP in an infrastructure IBFD WLAN:}
		\normalsize
		\setcounter{mytempeqncnt}{\value{equation}}
		\setcounter{equation}{16}
		\begin{align}
			&\mathcal{E}^{^{\text{AP}}}_{\text{d}}= \omega_{\text{d}}  \sigma\\
			&\mathcal{E}^{^{\text{AP}}}_{\text{\tiny S-TXRX}}= \omega_{\text{\tiny TX+CTRL}}   (\text{DATA}_{\text{\tiny TX}}+\text{ACK})+ \omega_{\text{\tiny RX+SIC}}  (\text{DATA}_{\text{\tiny RX}}+\text{ACK}) + \omega_{\text{d}}  (\text{DIFS+SIFS})\\
			&\mathcal{E}^{^{\text{AP}}}_{\text{\tiny C-TXRX}}= \omega_{\text{\tiny TX+CTRL}}  \text{DATA}_{\text{\tiny TX}}+ \omega_{\text{\tiny RX+SIC}}  \text{DATA}_{\text{\tiny RX}}+ \omega_{\text{d}} (\text{DIFS+SIFS+ACK})
		\end{align}
		\\
		\textit{State Probabilities for the AP in an infrastructure IBFD WLAN:}
		\begin{align}
			&Pr(\text{AP-d})= (1-\tau_{_{AP}})(1-\tau_{_{STA}})^{n-1}\\
			&Pr(\text{AP-S-TXRX})=\tau_{_{AP}}(1-\tau_{_{STA}})^{n-1}+(n-1)\tau_{_{STA}}(1-\tau_{_{STA}})^{n-2}\\
			&Pr(\text{AP-C-TXRX})=1-Pr(\text{AP-d})-Pr(\text{AP-S-TXRX})
		\end{align}
		\\
		\textit{Energy Consumption for a STA in an infrastructure IBFD WLAN:}
		\begin{align}
			&\mathcal{E}^{^{\text{STA}}}_{\text{d}} = \omega_{\text{d}}  \sigma\\
			&\mathcal{E}^{^{\text{STA}}}_{\text{\tiny S-TXRX}} = \omega_{\text{\tiny TX+SIC}}  (\text{DATA}_{\text{\tiny TX}}+\text{ACK})+ \omega_{\text{\tiny RX+CTRL}}  (\text{DATA}_{\text{\tiny RX}}+\text{ACK})+ \omega_{\text{d}}  (\text{DIFS+SIFS})\\
			&\mathcal{E}^{^{\text{STA}}}_{\text{\tiny S-$\overline{\text{RX}}$}} = \omega_{\text{\tiny RX+CTRL}}  (\text{DATA}_{\text{\tiny RX}}+\text{ACK})+ \omega_{\text{d}}  (\text{DIFS+SIFS})\\   
			&\mathcal{E}^{^{\text{STA}}}_{\text{\tiny C-TXRX}} = \omega_{\text{\tiny TX+SIC}}  \text{DATA}_{\text{\tiny TX}} + \omega_{\text{\tiny RX+CTRL}}  \text{DATA}_{\text{\tiny RX}} + \omega_{\text{d}}  (\text{DIFS+SIFS+ACK}) \\
			&\mathcal{E}^{^{\text{STA}}}_{\text{\tiny C-$\overline{\text{RX}}$}} = \omega_{\text{\tiny RX+CTRL}}  \text{DATA}_{\text{\tiny RX}} + \omega_{\text{d}}  (\text{DIFS+SIFS+ACK})
		\end{align}
		\\
		\textit{State Probabilities for an STA in an infrastructure IBFD WLAN:}
		\begin{align}
			&Pr(\text{STA-d}) = (1-\tau_{_{AP}})(1-\tau_{_{STA}})^{n-1}\\
			&Pr(\text{STA-S-TXRX})=\tau_{_{STA}}(1-p_{_{STA}})+(1-\tau_{_{STA}})^{n-1}\frac{\tau_{_{AP}}}{(n-1)}\\
			&Pr(\text{STA-S-$\overline{\text{RX}}$})=(n-2)\tau_{_{STA}}(1-\tau_{_{STA}})^{n-2}(1-\tau_{_{AP}})+\frac{(n-2)}{(n-1)}\tau_{_{AP}}\Big[\tau_{_{STA}}(1-\tau_{_{STA}})^{n-2}+(1-\tau_{_{STA}})^{n-1}\Big]\\
			&Pr(\text{STA-C-TXRX})=\tau_{_{STA}} p_{_{STA}}\\
			&Pr(\text{STA-C-$\overline{\text{RX}}$})=1-Pr(\text{STA-d})-Pr(\text{STA-S-TXRX})-Pr(\text{STA-S-$\overline{\text{RX}}$})-Pr(\text{STA-C-TXRX})
		\end{align}    
		\setcounter{equation}{\value{mytempeqncnt}}
		\hrulefill
		\vspace*{4pt}
	\end{figure*}
	
	\section{Results and Evaluation}
	Both analytical and simulation results are reported in this paper. First, power consumption and energy-efficiency are analyzed when the traffic is fully symmetrical (i.e. $\rho=1$) for both HD and IBFD systems. Then, the effect of symmetry on both power consumption and energy efficiency is presented through the two extreme cases of low symmetry ($\rho=0.1$) and high symmetry ($\rho=0.9$). The calculated power consumption values for $\omega_{\text{\tiny TX}}$, $\omega_{\text{\tiny RX}}$, and $\omega_{\text{d}}$ in TABLE \ref{murad.t2} are based on \cite{Lee15}. Values of $\omega_{\text{\tiny SIC}}$ and $\omega_{\text{\tiny CTRL}}$ are stated in \cite{Kobayashi18}. While $\omega_{\text{\tiny SIC}}$ accounts for both active and passive cancellation circuits, the majority of SIC is treated passively with minimal power consumption. 
	
	\subsection{Fully Symmetrical Traffic}
	Fig. \ref{murad1} shows how the number of nodes affects the power consumption per node in both HD and IBFD networks. Both analytical and simulation results are reported when the traffic is assumed to be fully symmetrical. This is the best case scenario where the link is fully utilized in both uplink and downlink directions. In the HD case, the AP and every STA have identical power consumption since HD IEEE 802.11 yields the same power profile for every node. The power consumption per node in an HD WLAN stabilizes at a constant value as the number of nodes increases since the dominant power consumption mode happens in states S-TX and C-TX, and the associated probabilities for both transmitting states reach a steady value quickly as the number of nodes increases. In the case of IBFD WLAN, the results are reported for both the AP and an STA since they have different properties here. Power consumption is higher in IBFD WLANs since there is simultaneous transmission and reception at both the AP and an STA when the channel is non-idle. The AP in an IBFD WLAN has high power consumption since it is always transmitting (states AP-S-TXRX and AP-C-TXRX) regardless of how many STAs are in the network. This does not constitute an efficiency concern since APs are typically powered by AC electricity in residential WLANs. As the number of nodes increases, power consumption per STA gradually decreases to reach a constant value since the probability values of transmitting states (i.e. STA-S-TXRX and STA-C-TXRX) quickly stabilize as the number of nodes becomes high. When $n=2$ with fully symmetrical traffic loads, the AP and the STA have the same power consumption in the IBFD case since they have the same probabilistic properties  with no collision and no overhearing from either nodes as indicated in \cite{Murad18}.
	\begin{figure}[!b]
		\centering
		\includegraphics[width=\columnwidth,trim=1.4in 3.3in 1.7in 3.5in,clip=true]
		{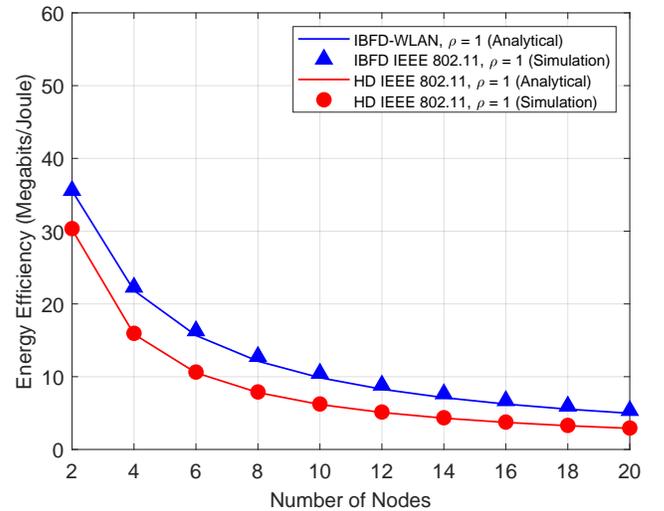}
		\caption{Energy-efficiency in HD and IBFD WLANs when $\rho=1$.}
		\label{murad2}
	\end{figure}
	
	Fig. \ref{murad2} shows analytical and simulation results of energy-efficiency in terms of Megabits/Joule resulting from dividing throughput by consumed power as in \cite{Kobayashi18}. The results are reported for both HD and IBFD cases. IBFD WLANs always have higher energy-efficiency since more data is transmitted. A key reason here is that only one node uses the link for data in HD networks while two transmitting nodes utilize the link for data in an IBFD WLAN.
	
	\subsection{High Symmetry vs. Low Symmetry}
\begin{figure}[!b]
		\centering
		\includegraphics[width=\columnwidth,trim=1.4in 3.3in 1.7in 3.5in,clip=true]
		{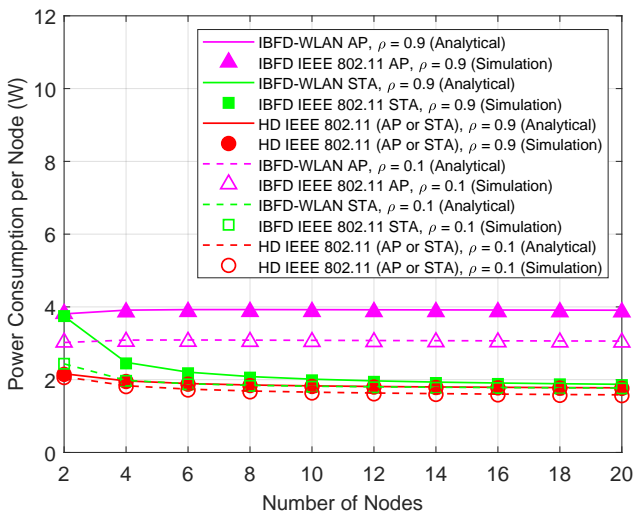}
		\caption{Power consumption per node in HD and IBFD WLANs.}
		\label{murad3}
	\end{figure}

	Fig. \ref{murad3} shows the results of power consumption per node in both HD and IBFD WLANs. High symmetry ($\rho=0.9$) and low symmetry ($\rho=0.1$) are considered. Similar patterns to the ones in Fig. \ref{murad1} can be seen here. Power consumption is reduced when symmetry is low since lower power consumption is needed to transmit (and receive) smaller uplink traffic loads. For HD WLAN, the power consumption is slightly affected by the change of symmetry mode but remains lower than the corresponding IBFD case. In the special case when an IBFD WLAN has two nodes, the network becomes more efficient due to the elimination of collisions \cite{Murad18}. In this case, the STA has higher power consumption than the corresponding HD case since it is simultaneously transmitting and receiving with IBFD. Power consumption increases as symmetry increases because more power is needed to transmit a larger uplink payload. The power consumption of the AP in the IBFD case with $n=2$ increases when the uplink load increases due to the increase in power consumption at the AP's receiver.
	
\begin{figure}[!t]
		\centering
		\includegraphics[width=\columnwidth,trim=1.4in 3.3in 1.7in 3.5in,clip=true]
		{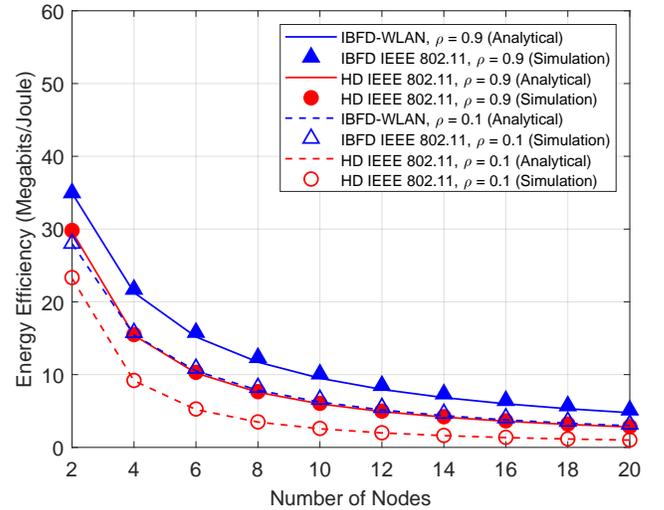}
		\caption{Energy-efficiency in HD and IBFD WLANs.}
		\label{murad4}
	\end{figure}

	Fig. \ref{murad4} shows the energy-efficiency for the cases of high and low symmetry scenarios in both HD and IBFD WLANs. The key result in this figure is that energy-efficiency of low symmetry IBFD WLAN is almost equal to the energy-efficiency of high symmetry HD WLAN. The low symmetry scenario is naturally inefficient due to the lower utilization of the uplink. Hence, it takes an extremely inefficient assumption (i.e. low symmetry) to reduce the high efficiency of an IBFD WLAN to the upper limit of energy-efficiency in the HD case. This result is intuitive since an IBFD WLAN with low symmetry has effectively only one fully utilized communications direction (i.e. downlink), which is equivalent to the fully utilized communications direction (i.e. either the uplink or downlink) in the HD WLAN with high symmetry. When $n=2$, the increase of energy-efficiency as the symmetry increases is due to the higher data amount transmitted over the link. Even though more power is needed when there is a larger data load, both HD and IBFD networks show increase in efficiency when the traffic is highly symmetrical. The increase of efficiency in an IBFD WLAN as symmetry increases shows that the increase of transmitted data is high enough to overcome the increase in consumed power.
	
	\section{Conclusion}
	An analytical model for power consumption and energy-efficiency in IBFD WLANs is presented and confirmed by simulation. Even though power consumption is higher in IBFD WLANs compared to HD WLANs, IBFD networks still have higher energy-efficiency. The presented model is necessary to study future IBFD solutions for IEEE 802.11 networks. 
	
	\ifCLASSOPTIONcaptionsoff
	\newpage
	\fi
	
	\bibliographystyle{IEEEtran.bst}
	
	

\end{document}